# Certification Grade Quantum Dot Luminescent Solar Concentrator Glazing with Optical Communication Capability for Connected Sustainable Architecture

Francesco Meinardi[1]*, Francesco Bruni[1], Claudio Castellan[2], Marco Meucci[3,4], Ali Muhammad Umair[4,3], Marcello La Rosa[1], Jacopo Catani[3,4]*, Sergio Brovelli[1]*

[1] *Dipartimento di Scienza dei Materiali, Università degli Studi Milano - Bicocca, via R. Cozzi 55, 20125 Milano, Italy.*

*2 Dipartimento di Fisica, Università degli Studi di Trento, via Sommarive 14, 38123 Povo (TN), Italy*

*3 Istituto Nazionale di Ottica del CNR (CNR-INO), via Nello Carrara 1, 50019 Sesto Fiorentino, Italy*

*4 European Laboratory for NonLinear Spectroscopy (LENS), via Nello Carrara 1, 50019 Sesto Fiorentino, Italy*

Corresponding author: francesco.meinardi@unimib.it, jacopo.catani@ino.cnr.it, sergio.brovelli@unimib.it

Energy sustainability and interconnectivity are the two main pillars on which cutting-edge architecture is based and require the realisation of energy and intelligent devices that can be fully integrated into buildings, capable of meeting stringent regulatory requirements and operating in real-world conditions. Luminescent solar concentrators, particularly those based on near-infrared emitting reabsorption-free quantum dots, are considered good candidates for the realisation of semi-transparent photovoltaic glazing, but despite important advances in optical property engineering strategies, studies of finished devices suitable for real-world operation are still lacking. In this paper, we demonstrate the first example of a fully assembled quantum dot luminescent solar concentrator-based photovoltaic glazing that meets all international standards for photovoltaic and building elements. We also show that these devices are capable of functioning as efficient Visible Light Communication (VLC) receivers even under full sunlight, thus combining energy and wireless connectivity functions in a realistic solution for smart, sustainable buildings.

## 1. INTRODUCTION

Given ongoing climate change and the serious geopolitical implications of energy supply, the transition to renewable energy is a priority that cannot be postponed. This urgency is now reinforced by EU[1] and other directives that require new buildings to be (nearly) zero energy and existing buildings to meet increasingly stringent energy requirements[2]. As a result, Building Integrated Photovoltaics (BIPV), which aims to integrate photovoltaic elements into the building envelope, plays a fundamental role, especially in urban contexts where there is little space for rooftop solar panel installation[3]. Exploiting the energy-passive space of the vertical facades of urban buildings is therefore a recurring motif in the development of new BIPV technologies. This includes both the realisation of opaque solar panels that are aesthetically compatible with urban design and, above all, the realisation of semi-transparent solar technologies that can replace the conventional glazing that makes up most of the surface area of contemporary buildings. In this context, luminescent solar concentrators (LSCs) offer the possibility of integrating photovoltaics in a way that is particularly well suited to architectural aesthetics[4], overcoming the common limitation of laminated photovoltaic glazing, which is characterised by varying degrees of opacity by the conventional PV panels or translucent thin films, which often dramatically worsens their aesthetic impact. In contrast, photovoltaic LSC glazing has no electrodes on its surfaces and offers the ability to select colour and degree of transparency to better meet the aesthetic and usability needs of buildings. This is due to the unique all-photonic operation of LSCs, which capture sunlight over large areas and concentrate it at their edges in the form of optically guided luminescence within transparent waveguides (**Figure 1a**)[5]. The opaque PV elements that perform the optical-electrical conversion are ultimately hidden within the window frame, resulting in photovoltaic windows that are seemingly indistinguishable from conventional semi-transparent ones. Research over the years has addressed a number of material design and manufacturing challenges[6], including the suppression of self-absorption by the chromophores embedded in LSC waveguides, which has prevented the fabrication of large LSC devices for many decades, and the extension of solar coverage into the near infrared, which increases PV efficiency and imparts a neutral colour suitable for residential glazing. A major turning point in LSC technology was the introduction of colloidal semiconductor quantum dots (QDs) as wide Stokes shift emitters[4c, 7]. To date, the best combination of performance and aesthetics has been achieved by LSCs based on QDs of group I-III-VI$_2$ ternary semiconductors (e.g. $CuInS_2$/ZnS, $AgInS_2$/ZnS, etc.), which exhibit a broad featureless absorption spectrum in the Vis-NIR region and efficient Stokes-shifted luminescence in the NIR[7e, 8]. Significant progress has also been made in the fabrication of plastic waveguides of high optical quality using industrial methods suitable for mass production[6, 8a], and in the consolidation of guidelines shared by the scientific community for the photovoltaic characterisation of devices[9]. Thanks to the tremendous efforts of the QD and LSC communities, LSC devices are approaching single module efficiencies close to real-world applications (between 1-2% for devices of ~400 cm$^2$), and a further substantial increase in power output is expected once residual loss processes, such as partial reabsorption by QDs or polymer waveguides and outcoupling losses to peripheral cells, are minimised. It is important to note that, even at current levels of performance, the total power output generated by LSC glazed surfaces of high-rise buildings makes a significant contribution to energy demand due to the huge multiplication factor offered by the available vertical area. In addition, and no less importantly, the power generated in situ by LSC glazing is already sufficient to support smart functions of the glazing itself, effectively becoming a self-powered integrated device for interconnectivity, sensing and the Internet of Things (IoT).

However, although there has been considerable progress in the development of LSC emitters and waveguides, there have been very few studies of LSC prototypes in an operational environment[10], most notably the excellent work by Debije and co-workers on LSC-based solar noise barriers[11], and no demonstration of fully assembled LSC-based insulating glass units (LSC-IGUs) that can be effectively implemented in residential buildings. These take the form of triple IGUs, modified by replacing the inner pane with the LSC waveguide (which is thus protected from external agents), separated from the outer panes by sealed spacers that create two distinct insulating chambers that provide thermal insulation and prevent moisture ingress. Crucially, since the LSC-IGU is both a PV device and a building component, it has to comply with very strict regulations in both areas, as set out in its respective international certifications. Passing these certifications will provide LSC-based BIPV devices with an unprecedented high level of integrability that is suitable for both new build and for the sustainability-driven renovation of existing and/or historic buildings. Crucially, the benefits of LSC integration go beyond energy aspects and offer the possibility of advancing a second fundamental pillar of contemporary architecture, namely connectivity and IoT, by realising self-powered, networked intelligent facades capable of exchanging information and receiving instructions for specific functions (electrochromic colouring; electrocatalytic air purification, integrated sensors, wireless data transfer, etc.) via modulated visible light harvested and concentrated by the LSC panel. This aspect is crucial in the 6$^{th}$-generation global network architecture (6G), *ab-initio* encompassing the optical domain to deploy fast and pervasive wireless links among the exponentially increasing number of connected devices, with reduced latencies and low energy impact[12]. In this context, we have recently shown that $CuInS_2$ QD-based LSC waveguides function as antennas for long-distance (60 m) error-free optical wireless communication (OWC) at baud rates up to 1 Mb/s using modulated 405 nm light[13]. In order to progress in this direction, it is, however, necessary to demonstrate their capability to work in real operating conditions for visible light communications (VLC) and light-fidelity (Li-Fi)[14] using modulated standard indoor white light sources under full solar irradiance.

In this work, we aim to contribute to this challenge by demonstrating the first example of a complete LSC-IGU capable of passing the most severe tests required by the international certifications for PV modules and glazing systems respectively, which are necessary for the installation in real built contexts of PV modules or IGUs, featuring white-light VLC and Li-Fi connectivity as optical receiver. To this end, fully assembled LSC-IGU prototypes have been designed and manufactured in an automated industrial process using state-of-the-art LSCs based on $CuInS_2$ QDs embedded in mass-polymerised polymethylmethacrylate (PMMA) sheets. Remarkably, the devices showed perfect stability for over one year of continuous outdoor testing and accelerated ageing tests involving low/high temperature cycles, intense UV irradiation and exposure to moisture were successfully passed without any degradation in photovoltaic activity or structural integrity. Luminous and thermal transmittance evaluated according to ISO standards also after certified UV stress tests further demonstrate that the LSC-IGUs function as an effective insulating envelope with performance comparable to commercial triple-glass IGUs. No less important, to explore one of the most relevant opportunities of VLC technology in the IoT scenario, which is the realization of hybrid energy harvesting and optical communication device[15], we characterize the performances of the LSC slab, contained in the LSC-IGUs, as optical receiver in white-light VLC applications. By integrating a custom photoreceiver on one edge of the panels used in our LSC-IGUs, and using a modulated white light LED-based spot lamp (common for artificial lighting in homes, retail and public offices) as the transmitter, we demonstrate the ability of the LSC-IGU to establish a VLC link in a relevant scenario, assessing the range and quality of communication for realistic lamp-to-window positions and under strong solar irradiation. Our work exploits the opportunities unleashed by VLC and Li-Fi in the IoT revolution, paving the way for the realisation of hybrid smart windows with combined energy harvesting and VLC communication capabilities.

## 2 Results and Discussion

*2.1 QDLSC-IGU fabrication, assembly and outdoor durability.* The LSC-IGUs with $40.5 \times 40.5$ cm$^2$ size were assembled using state-of-the-art PMMA LSCs (thickness = 0.7 cm, see top left photograph in **Figure 1b**) produced via industrial radical mass polymerization of MMA mixtures containing 0.05%wt of CuInS$_2$/ZnS QDs (see Methods for synthesis details) resulting in optical absorptance of 70% of the solar spectrum in the 400-700 nm range according to the international standard EN 410: 2011[16]. The optical absorption and photoluminescence (PL) spectra of the LSCs are reported in **Figure 1c** showing the typical broadband seemingly featureless absorption profile of CuInS$_2$/ZnS QDs and a largely Stokes shifted PL peak at ~780 nm. The PL quantum yield of the QDs inside the LSC was 70±5%. Consistent with previous results, the polymerization procedure ensured fine dispersion of the QDs inside the polymer matrix, resulting in residual absorption background at the PL wavelength matching the intrinsic absorption of pure PMMA (inset of **Figure 1c**)[8a]. The absorption peak at 900 nm was due to the C-H stretching overtone and set the long wavelength limit of the transparency window of the LSC waveguide, whereas the short wavelength end was determined by the absorption edge of the QDs at ca. 700 nm. As a result of suppressed reabsorption and negligible scattering losses, the LSC waveguide behaved essentially ideally, as shown by the good agreement between the experimental data collected using a calibrated 1.5 AMG solar simulator for a 20 cm × 20 cm slab (corresponding to the maximum area illuminated by the simulator) and the Monte Carlo simulated relative power vs. illuminated device area calculated for an ideal reabsorption/scattering free LSC of identical dimensions (**Figure 2d**). Once the quality of the QD-PMMA waveguide was established, we proceeded to assemble the LSC-IGU shown in **Figure 1e** according to the scheme sketched in the same figure, using the custom industrial equipment shown in **Figure 1a**. Specifically, LSC panels were laser cut to size and edge polished. Customised 98 mm × 11 mm c-Si mini-modules, consisting of ten 9.8 mm × 8 mm cells mounted in series, were connected in parallel to form continuous PV strips of electrically independent modules of the same length as the LSC edge. The PV strips were then coupled to the LSC waveguide by in-situ polymerisation of a polyacrylate binder, which ensured mechanical robustness and a continuous refractive index across the LSC/PV cell interface. Finally, to integrate the LSC into an IGU (size 43.8 × 43.8 cm$^2$) consisting of internal and external glass plates and thermally insulating spacers, the edges of the device (about 1.5 cm) were encapsulated in metal cable trays specially designed with a thin notch and a compressible filler to ensure minimal physical contact with the LSC waveguide - to minimise optical losses - and to compensate for the different thermal expansion of PMMA (70 ppm/°C) with respect to the borosilicate glass (~10 ppm/°C) forming the external plates (**Figure 1f**). Further details on the IGU assembly procedure are reported in Methods. The production cost of the prototypes were approximately $500/m$^2$, making them competitive with energy passive triple glazing. Comparison with existing BIPV technologies is difficult because the cost of PV in general depends dramatically on economies of scale, which do not exist for LSC-IGUs but will realistically reduce the cost of commercial products. The external efficiency of the LSC at the various stages of assembly was characterised under outdoor conditions using a calibrated silicon PV module with identical orientation as a reference and holding the LSC more than 1 m above the ground to avoid back reflection effects of transmitted sunlight. As shown in **Figure 1g**, the power conversion efficiency of the LSC was PCE=0.9%, which decreased to 0.72% with the addition of the cable tray, indicating that despite the bespoke design, some optical losses still occurred along the edges of the device. The current limiting factors for the PCE of LSC-IGUs are therefore to be found both in the optical properties of the QDs and in the device assembly. In particular, the major detriment for the LSC optical output is due to the residual reabsorption by polymer matrix and by the broad low-energy absorption tail of the QD of the corresponding luminescence that also presents itself as a relatively broad peak, which increases both the

geometrical and nonradiative losses due to non-unit emission yield. On the device assembly front, beside the above-mentioned effect of the cable tray, substantial losses in electron energy are due to the low open circuit voltage of the c-Si PV cells ($V_{OC} \sim 0.56$ V) which call for alternative solutions (e.g. InAs) that are however not yet commercially viable for large scale LSC-IGU production. Incorporation into the final IGU did not alter the power efficiency and ensured perfect stability in operating condition for over a year of continuous outdoor monitoring. This is highlighted in **Figure 1h**, where we show the photocurrent extracted from half of the LSC perimeter (the other half was kept as a backup in case of device failure), over a period of seven days between 10 July and 17 July 2021 and 2022, together with the corresponding temperature measured with a thermal sensor coupled to the LSC waveguide. Remarkably, the LSC-IGU showed perfect stability over time and good agreement between LSC power and device temperature on each day tested. Note that during the whole measurement period of over one year the IGU has not been treated in any way (not even cleaned).

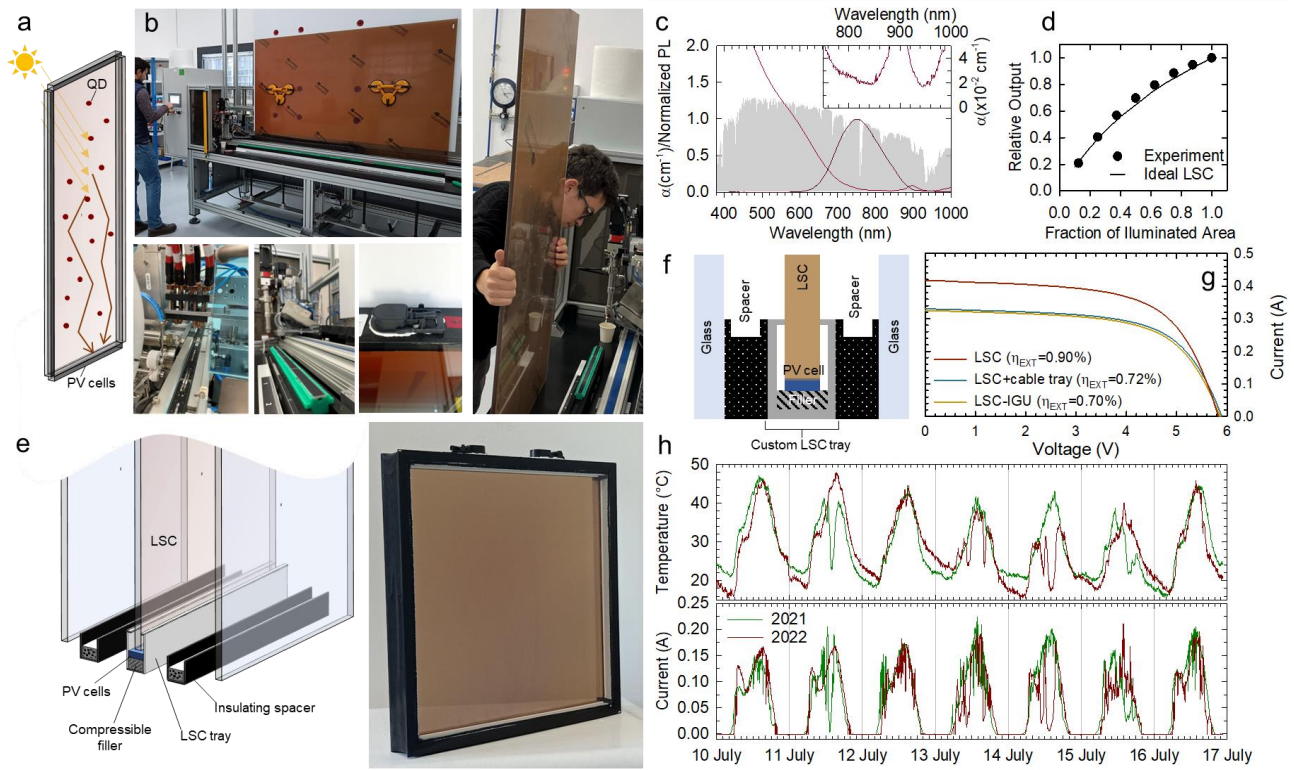

**Figure 1. Assembly and operational stability of IGUs based on QD-LSCs. a.** Schematic of the functioning mechanism of an LSC under solar irradiance. **b.** Photographs of the fabrication and assembly line for LSC-IGUs including cutting and polishing, assembly and installation of the perimeter c-Si mini-modules and detail of the Twin Box electrical connector at the edge of the IGU. **c.** Optical absorption and photoluminescence spectra of a QD-LSC. Inset: detail of the absorption profile in the 750-1000 nm range highlighting minimal background due to reflection and no additional scattering losses. **d.** Normalized relative output of the IGU-LSC (20 cm × 20 cm slab, circles) under simulated sunlight compared to the Monte Carlo calculated response of an ideal scattering/reabsorption-free LSC. **e.** Exploded view and photograph of the complete LSC-IGU. **f.** Sketch of the side view of the LSC-IGU highlighting the custom metal cable tray and spacers. **g.** Current-voltage characteristic measured in outdoor conditions for a QD-LSC (30 × 30 cm slab) without (brown curve) and with the cable tray (green curve) as well as incorporated in the IGU (yellow line). **h.** Internal temperature and current output of the same LSC-IGU measured over seven days in July 2021 and 2022 showing identical behaviour. The device has been kept outdoors for the whole year.

**2.2 Photovoltaic certification of QDLSC-IGUs.** Crucial to the real-world application of LSC glazing systems in BIPV is the ability of the LSC-IGU to comply with international regulations for both PV modules and building components. With this in mind, we have subjected two identical LSC-IGUs to some of the most stringent tests foreseen by the IEC 61215-1:2021 standard[17], which defines the requirements for the design qualification of terrestrial PV modules suitable for long-term operation in open air climates. The tests include an initial preconditioning with intense UV light (228 hours for a total irradiation of 15.1 kWh/m² for sample 1; 900 hours for a total irradiation of 60.1 kWh/m² for sample 2, both at a constant temperature of 60±5°C, **Figure 2a**) followed by either thermal cycling combined with humidity freeze tests (sample 1) or damp heat tests (sample 2). The respective temperature ramps are shown in **Figure 2b** and **2c** and, in particular, for the thermal cycles, consisted of 50 full cycles between -40°C and +85°C starting from an initial temperature of 25°C, with a maximum heating/cooling rate of 100°C/h and minimum dwell times of 10 minutes; for the humidity freeze test, in ten full cycles consisting of 20 hours at 85°C and 85% relative humidity (RH) followed by 30 minutes at -40°C with a cooling rate of 100°C/h (between +85 and 0°C) and 200°C/h (between 0 and -40°C). For the damp heat test, the LSC-IGUs were heated to 85°C (maximum heating rate 100°C/h) and kept at this temperature and 85% RH for 200 hours. We stress that these tests have been chosen by international committees to simulate ageing of standard opaque PV modules that reach temperatures above 80 °C. On the other hand, the operating temperatures of semi-transparent LSC-IGUs are much lower (maximum temperature in July is below 50°C, **Figure 1h**). Therefore, in contrast to standard photovoltaic modules, for LSC-IGUs the tests conditions required by the norm are much more severe than the actual operating situations. In accordance with the IEC 61215-1:2021 regulation, the PV characteristics of the LSC-IGUs were assessed using a calibrated flash test with simulated sunlight, which is commonly used to measure the output power compliance of a solar PV module. The results of the test are summarised in **Figure 2d-f** and show that the I-V characteristics and the relative efficiency were not affected by the harsh conditions imposed by both tests as well as the UV preconditioning. The slight increase in efficiency after the thermal cycling and humidity freeze tests could be due to improved optical coupling between the LCS waveguide and the perimeter PV cells caused by the high temperature annealing of the polymeric adhesive. Visual inspection of the units after the tests also revealed no significant variation in the structure of the IGU or the internal LSC. Most notably, these results demonstrate that IGUs containing QD-LSCs meet the standard certifications required for commercial PV modules, an important milestone for this technology.

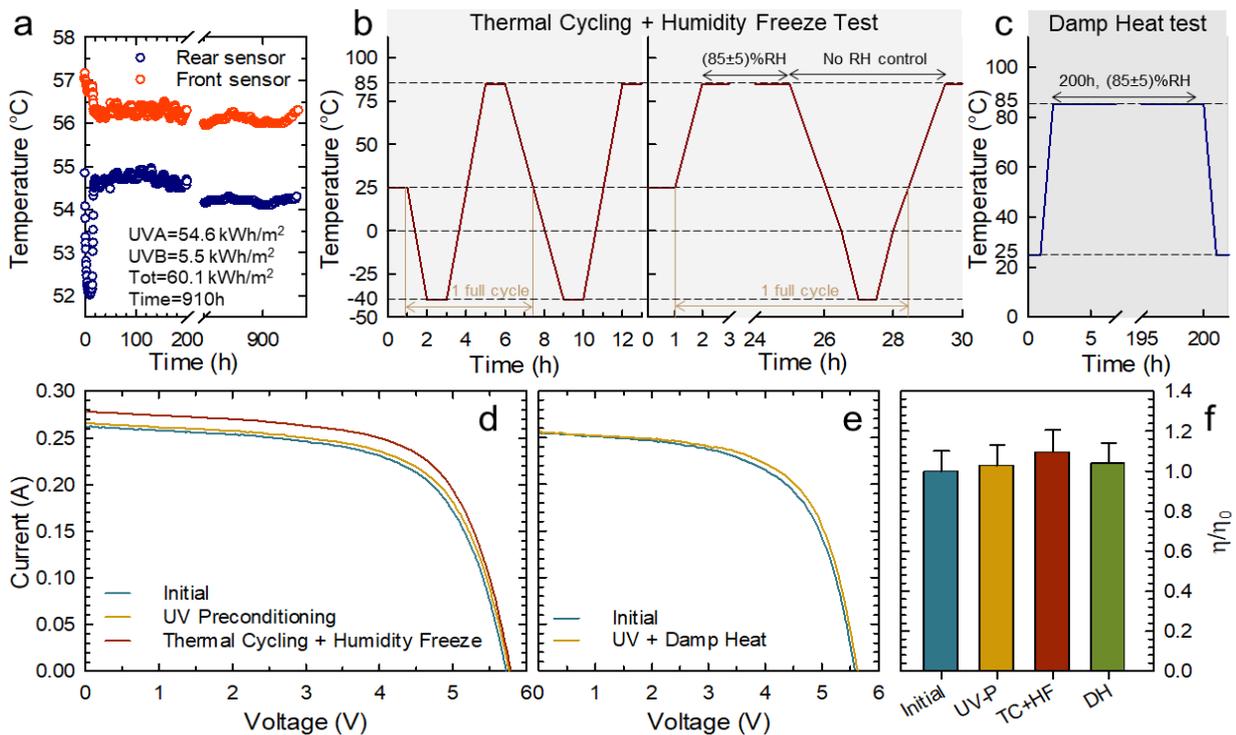

**Figure 2. Stress test according to IEC 61215-1:2021. a.** Internal temperature during preconditioning with intense UV light. Sketches of the temperature and humidity ramps of **b.** the thermal cycling and humidity test and **c.** the damp heat test required by the standard. Current-voltage characteristics of the LSC-IGU at the different stages of **d.** the thermal cycling and humidity test and **e.** the humid heat test. **f.** Summary of the relative performances normalised to the initial conditions after all the tests.

*2.3 QDLSC-IGUs as building elements.* Having demonstrated the ability of IGUs embedding QD-LSC to comply with standard photovoltaic certification, we focused on assessing their suitability as construction elements by performing the tests and calculations required for glass in buildings. These included moisture penetration tests (EN 1279-2:2018)[18], light transmission assessments (EN 410: 2011)[16] and thermal transmittance calculations according to ISO 10077-1:2107[19] and ISO 12631:2017[20]. The moisture penetration tests provided for in EN 1279-2:2018 consist of evaluating the percentage of water content (indicated as $W$) in molecular sieves (Siliporite® NK 30, 3Å from Arkema) placed inside the sealed chamber of IGU prototypes before and after 11 weeks of exposure to heat and moisture according to the scheme shown in **Figure 3a**, specifically consisting of 56 thermal cycles (-18/+53°C) at RH>95% for a total period of 4 weeks, followed by 7 weeks at a constant temperature of 58°C and RH>95%. The $W$-value is measured at the beginning ($W_i$) and at the end ($W_f$) of the test. Given the moisture adsorption capacity of desiccant $W_C$ (in our case we measured 16.2%) it is possible to determine the moisture penetration index $I = (W_f - W_i)/(W_C - W_f)$. The test is passed if, at the end of the full ramp, the average $I$-value between five different prototypes is $I_{AV}<20\%$ and no prototype has $I>25\%$. In our case the average initial water content was $W_{i,AV}$=2.38%, meaning that the average final water contents had to be $W_{f,AV}<5.1\%$ and no prototype should show $W_f>5.8\%$. In addition to the standard protocol, which only requires the initial and final measurements, we also monitored the water content during the ramp to identify any trends in moisture ingress into the LSC-IGUs. Since these are destructive testing, we produced 35 identical LSC-IGUs to perform all these measurements and we monitored them at different stages of the ramp. As shown in **Figure 3b,** $W$ increased linearly with test time, reaching $W_{AV}$=5.0% (corresponding to $I_{AV}$=18.9%), with the highest value of 5.49% (corresponding to $I$=20.7%), well within the test requirements. Visual inspection of the IGU after the test, which is also required by the standard, also showed no fogging from internal moisture (**Figure 3c**). Overall, these results demonstrated that LSC-IGUs also comply with the EN 1279-2:2018 standard[18]. Nevertheless, the penetration of some moisture into the IGU raised the question of whether the sealants and cable trays used to incorporate the LSC, or the electrical wiring to the electrical connector at the edge of the IGU, were responsible for the moisture penetration during the test, or whether it was caused by the release of adsorbed water by the PMMA LSC panel during heating. To address this issue, we performed the same test on LSC IGUs without the wiring and on triple IGUs with identical sealants and spacers but without the LSC waveguide. As shown in **Figure 3b**, the LSC without wiring (squares) showed similar behaviour to the complete LSC IGUs, whereas the removal of the LSC panel resulted in very low $W$ values, indicating that the source of moisture was likely to be the LSC panel, which is manufactured by the cell casting process. This is also confirmed by the lower moisture level measured in the IGUs without LSC (measured at the end of ageing) than that of LSC-IGUs (measured before ageing), probably resulting from released of moisture by the LSC in the period of about one month between the IGU fabrication and the tests. This further suggested that, although not required for certification, the water penetration index of LSC-IGUs could potentially be further reduced by post-fabrication annealing and drying of the waveguides.

The luminous, solar, and thermal characteristics of IGUs are crucial for their integration into buildings as they contribute to determine lighting, heating and cooling requirements and permit comparison between different types of envelope technologies. We therefore evaluated the luminous transmittance of our QD-LSC waveguides according to the EN 410:2011 [16] and ASTM C1371-04a[21] standards. A photograph of the LSC-IGU in outdoor conditions is shown in **Figure 3d**. For quantitative assessments of the optical properties of the device the regular spectral transmittance at normal incidence and spectral reflectance at an angle of incidence of 5° in the 300-2500 nm range were measured using a dual-beam spectrophotometer with an aperture angle of 2°, providing near-parallel and near-normal incident radiation. A 6 mm thick Suprasil W ultra-pure silica glass was used as a reference for reflectance measurements. The photometric accuracy of the spectrophotometer in the visible range was checked using certified filters (SRM 930D). The light transmittance, light reflectance, direct solar transmittance, direct solar reflectance and solar factor of the glazing were then determined according to the calculation procedures specified in EN 410:2011. The values are listed in **Figure 2e**.

Although not defined by a specific standard, another parameter of great importance for a light source is the colour rendering index (CRI), which is a quantitative measure of its ability to faithfully reproduce the colour of objects compared to an ideal or natural light source. We thus calculated the CRI of our LSCs as if they were filters for standard illuminant D65, using the CIE13.3 method and eight Munsell test colour samples. In the case of an LSC with 40% transparency (calculated according to EN 410), as in the case of the LSC-IGU shown in **Figure 3c**, the CRI obtained is 96, which places it in the best category (1A) defined by UNI 10380 and recommended for the lighting of homes, museums, graphic studios, hospitals, etc. Even when the degree of transparency is varied, the CRI always remains very high, and only when the transparency is reduced to 10% does the CRI fall slightly below 90 (see Table S1 in the Supplementary Information for the complete dataset).

The hemispherical surface emissivity, $E$, of the QD-LSC at room temperature was assessed in accordance with ASTM C1371-04a protocol using a calibrated emissometer interfaced with a scaling digital voltmeter, yielding $E$=0.91, which is comparable to common glass (~0.9). The solar radiation resistance of the spectral characteristics was then evaluated according to the UNI EN ISO 12543-4:2022 standard[22] by exposing 3 QD-LSCs to 16 OSRAM ULTRAVITALUX 300 W lamps for 2000 hours at 45±5 °C, as specified in paragraph 7.3.1 of the standard (Method A). Before and after irradiation, the light transmittance was measured in accordance with the EN 410:2011 standard already used to evaluate the optical characteristics of the LSC-IGUs shown in **Figure 3d**. The requirement of the UNI EN ISO 12543-2: 2022 standard (in point 5.3) is that the transmittance of the three samples irradiated should not vary by more than ± 2% of their absolute initial value. As can be seen from the data in **Figure 3f**, all the devices showed excellent durability under solar irradiation with a maximum change of only 1.6%, again a remarkable result never reported before for QDs-LSCs and an important step towards their architectural integration. Finally, we calculated the thermal transmittance of our complete LSC-IGU for window ($U_W$) applications following the protocols described in ISO 10077-2:2017 standard[19-20].

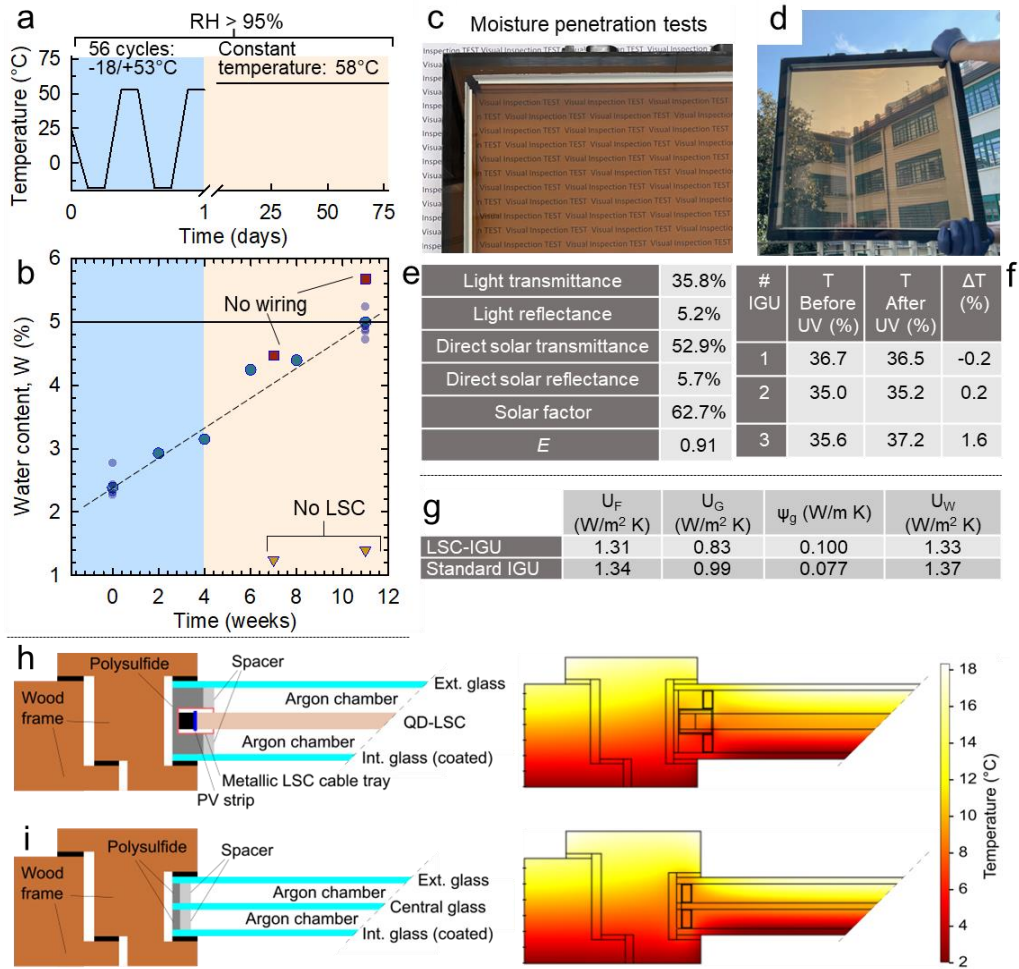

**Figure 3. Evaluation of the LSC-IGU as construction element. a.** Sketch of the temperature and humidity ramps of the moisture penetration tests (EN 1279-2:2018)[18]. **b.** Water content vs. time for a fully assembled LSC-IGU (blue circles), for an LSC-IGU without the electrical wiring (purple squares) and for a triple IGU without the LSC panel (triangles). **c**. Photograph of the LSC-IGU after the moisture penetration test showing no fogging due to water condensation. **d**. Photograph of the LSC-IGU in ambient illumination outdoors highlighting the transparency and colour neutrality. Results of **e.** the light transmission assessments (EN 410: 2011)[16], **f.** the solar radiation resistance (UNI EN ISO 12543-4:2022)[22] and **g.** the thermal transmittance calculations according to ISO 10077-1:2107[19] and ISO 12631:2017[20]. **h,i** Sketches of the device configurations considered for the thermal transmittance calculations (top panel: LSC-IGU, bottom panel: conventional triple IGU) and respective temperature distributions computed considering an inside-to-outside thermal difference of 20°C.

The estimation of $U_W$ refers to the specific pane configuration shown in **Figure 3g**, which we compared to a conventional IGU shown in **Figure 3h**. The main results are listed in **Figure 3i**. According to the ISO 10077-1:2017 standardization, $U_W$ depends on the frame-glass junction thermal transmittance ($\Psi_g$, referred to as the *window perimeter* parameter), the frame thermal transmittance ($U_F$), and the glazing thermal transmittance ($U_G$, referred to as the *window area* parameter), which depends on the composition of the IGU panes and was calculated according to the EU standard EN 673:2011. For the LSC-IGU, we found $U_G$=0.83 W/m²K, slightly lower than the conventional IGU with $U_G$=0.99 W/m²K. For the frame material, we considered wood in both cases and obtained $U_F$=1.31 W/m²K and 1.34 W/m²K respectively. The same approach was used to calculate $\Psi_g$, yielding $\Psi_g$ = 0.100 W/mK for the LSC-IGU, which was slightly higher than the 0.077 W/mK value for the standard IGU, due to the presence of the stainless-steel cable tray used to contain the LSC panel edge and the PV cells. Applying an inside-to-outside thermal difference of 20°C, we obtained the temperature distributions shown in the right-hand panels of **Figure 3g** and **3h**. Finally, combining all these parameters we evaluated the global thermal transmittance $U_W$. For a 1×1 m² LSC-IGU, we obtained $U_W$=1.33 W/m² K, which is slightly lower than for a standard IGU with $U_W$=1.37 W/m². This indicates that LSC-IGUs are suitable as thermal insulation envelope elements, with no disadvantage compared to conventional energy-passive technologies due to their additional PV function.

*2.4 LSC-IGU for wireless communications.* To explore the feasibility of hybrid, smart windows with energy harvesting and VLC communication capabilities based on our LSC-IGUs, we experimentally investigate the performances of the embedded LSC slab as optical receiver (RX) for VLC purposes, and in particular in Li-Fi applications exploiting white light LED sources as transmitters (TXs). In this context, we implement a VLC transmission system composed of a high-power white spotlight type COB-LED lights (Exenia Museo Mini 2L, 300 KHz modulation BW) as VLC TX stage, positioned on the ceiling (approx. 3 m from the floor) in order to reproduce a realistic scenario (**Figure 4b,c**). On the RX side (sketched in **Figure 4d** as teal box), we use a QD-LSC slab with dimensions of 38 cm × 38 cm from the same production lot as those employed in the fully assembled IGUs discussed above. The slab is coupled with a rectangular photodiode (PD) with an active area of 3 mm × 30 mm (Hamamatsu, S3588-09) placed in correspondence of the middle point of one of the slab's edges (**Figure 4a**). The PD has dimensions approximately equal to those of the solar cells used in the standard use of LSC-IGUs, and less than a PV module should be removed to allocate the PD on the edge of the LSC. At this proof-of-concept stage, the LSCs with PD were not fully integrated into the IGUs as not all components were miniaturized to be compatible with the assembly equipment. However, under the conditions used, we do not expect the solar energy collection efficiency to decrease significantly after the insertion of the VLC receiver element. The photocurrent generated by PD is amplified by a custom, AC-coupled transimpedance amplifier (TIA) stage. The received signal after TIA is digitised by means of a Schmitt trigger comparator, decoded and byte-wise compared with a preloaded reference packet for *Packet Error Rate (PER)* calculation by a digital microcontroller board (Arduino DUE). The same board is used to implement the VLC TX stage (**Figure 4d**, pink box). The TX stage continuously transmits packets of 32 bits, inserting digital information as intensity modulation into the optical carrier emitted by the spotlight, through a suitably designed current driver. In our VLC tests, the LED spotlight is fixed on the ceiling pointing downwards, at a height $z$ = 2.8 m, corresponding to an average height for a common room, in order to reproduce a realistic scenario. The slab is positioned at $z$ = 1.55 m from the floor, and the coordinates $x$ and $y$ are varied. This choice simulates different reciprocal positions between a Li-Fi source and a smart window enabled OWC. We measure the received optical signal as a function of $x$ and $y$, reporting data in **Figure 4e**. The contour map is obtained by applying a triangulation algorithm on the experimental data, identified by the black dots. The lower part of the same panel shows SNR values measured on the central line ($y = 0$) as $x$ varies. For SNR calculation we consider the ratio between measured amplitude on each point and the measured RMS noise value. The received signal mainly depends on the photometric characteristics of the LED spotlight and the mutual position between the slab and the light source. We observe that the largest SNR values are recorded in the region (0 m < $x$ < 1 m; -0.5 m < $y$ < 0.5 m). At short distances along $x$ ($x$ < 0.5 m), the light hits the slab in a very grazing way, reducing the effective surface hit by radiation and hence lowering the collected signal despite the large intensity. For large y distances, instead, the reduction is mainly due to the vanishing intensity inherent to the typical emission lobe of the LED source. To assess the quality of the optical channel, the experimental tests consist in measuring *PER*, defined as the ratio between the lost packets compared to the total number of transmitted packets. A packet is considered wrong if a single bit is lost during transmission. The lower observable PER ($10^{-5}$) is limited by the maximum number of transmitted packets ($10^5$). **Figure 4f** shows the PER values measured as the distance x and y vary. As in the previous **Figure 4e,** the contour map is obtained applying the same algorithm on the experimental data (black dots). *Error-free* transmission is considered for PER values lower than $10^{-5}$ (light green area). Below the map is shown the PER vs *x* (m) measured on the central line for *y* = 0 m. The shaded red line shows the calculated PER on the transmission considering the relation between SNR and Q-function as reported in ref.[13].

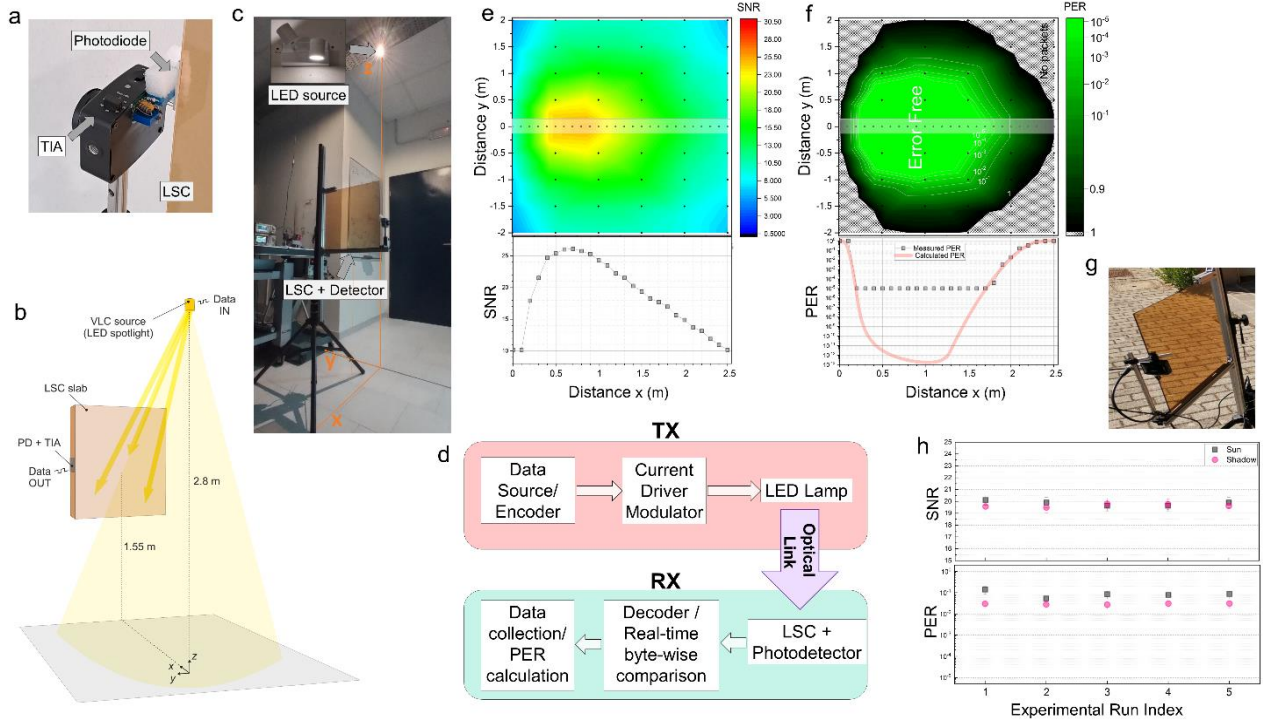

**Figure 4 QD-LSC-IGUs as optical antennas for VLC communication. a.** Custom photoreceiver (PD + TIA) coupled to LSC edge. **b, c).** Sketch of the VLC test setup and coordinate system used. **d**. TX and RX block diagram. **e.** SNR measurements and, **f**, PER measurements for different mutual positioning between spotlight and LSC. **h**. LSC slab outdoors. **i**. Repeated VLC runs, in "*sun*" (black) and "*shadow*" (red) conditions.

As in the previous **Figure 4e**, it can be observed that transmission cannot be established for very grazing distances, in which the positioning between the LED spot and the slab forms a very large angle (grey area on the contour map). In view of assessing the feasibility of hybrid, smart windows capable of joint solar harvesting and VLC based on our LSC-IGUs, we also test their OWC performances in conditions of high solar irradiance. Indeed, smart, hybrid devices must support reliable Li-Fi and/or VLC links in presence of strong illumination background, which could severely affect the performances of a VLC system because of strong saturation effects in the receiving stage. The LSC slab is positioned outdoors under direct solar illumination (**Figure 4g**), and the reciprocal position between spotlights and LSC is made in such a way as to reproduce SNR approx. 20, corresponding to PER = $10^{-2}$ measured in laboratory conditions. **Figure 4h** shows a comparison between direct sunlight illumination ("*Sun*"), corresponding to illuminance level of more than 70000 lux, condition which represents the case in which the window is hit directly by the sun, and no direct illumination ("*Shadow*"), corresponding to ~10000 lux, which represents a daytime scenario where the window is not directly hit by the sun. We observe a complete equivalence, within experimental errors, between the communication performances in the two conditions, for five repeated runs extending over different days.

## 3 Conclusion

In summary, we have realised the first example of photovoltaic triple glazing units based on QD-LSCs and tested them to the international standard for both BIPV and architectural glazing. Our results show that these LSC-IGU prototypes meet all the stringent requirements of the international regulations in both areas, and thus represent a valuable and realistic technology for the integration of renewable energy into the building envelope. The current PCE of LSC-IGUs is largely sufficient to self-power integrated technologies, such as the VLC capability, as well as other more established domotics functions such as environmental sensing, electrochromism, etc., making this technology particularly adaptable for retrofitting existing buildings without additional wiring. In order to extend the functionality of our LSC-IGUs for networked 'smart' architecture, we have therefore investigated their suitability as optical antennas for light communication and demonstrated their remarkable ability to operate as reliable Li-Fi and VLC links under strong sunlight illumination operating conditions. Our results represent an important step towards the practical realisation of self-powered smart buildings.

**Acknowledgements**

This work was funded by the Horizon Europe EIC Pathfinder programme, project 101098649 - UNICORN and by the Italian Ministry of Research PRIN programme, project IRONSIDE.

**Competing interests**

The authors declare no competing interests.

**References**

[1] European Commission, *Official Journal of the European Union, Directive (EU) 2018/2001 of the European Parliament and of the Council on the promotion of the use of energy from renewable sources* **11 December 2018**.

[2] a) European Commission, *Official Journal of the European Union, Directive 2010/31/EU of the European Parliament and of the Council on the energy performance of buildings* **19 May 2010**; b) European Commission, *Communication from the Comission* **2015**, C/2015/6317.

[3] a) D. D'Agostino, S. T. Tzeiranaki, P. Zangheri, P. Bertoldi, *Energy Strategy Rev.* **2021**, 36, 100680; b) H. Gholami, H. Nils Røstvik, K. Steemers, *Energies* **2021**, 14, 6015; c) P. Macé, A. E. Gammal, D. Mueller, H. Bürckstümmer, presented at The European Photovoltaic Solar Energy Conference and Exhibition (EU PVSEC) **2016**; d) M. Vasiliev, M. Nur-E-Alam, K. Alameh, *Energies* **2019**, 12, 1080.

[4] a) Y. Zhao, R. R. Lunt, *Adv. Energ. Mater.* **2013**, 3, 1143; b) B. S. Richards, I. A. Howard, *Energy Environ. Sci.* **2023**, 16, 3214; c) F. Meinardi, Q. A. Akkerman, F. Bruni, S. Park, M. Mauri, Z. Dang, L. Manna, S. Brovelli, *ACS Energy Lett.* **2017**, 2, 2368; d) Y. Nie, W. He, X. Liu, Z. Hu, H. Yu, H. Liu, *Building Simulation* **2022**, 15, 1789; e) G. Yu, H. Yang, D. Luo, X. Cheng, M. K. Ansah, *Renew. Sust. Energ. Rev.* **2021**, 149, 111355.

[5] F. Meinardi, F. Bruni, S. Brovelli, *Nat. Rev. Mat.* **2017**, 2, 17072.

[6] S. Castelletto, A. Boretti, *Nano Energy* **2023**, 109, 108269.

[7] a) F. Meinardi, A. Colombo, K. A. Velizhanin, R. Simonutti, M. Lorenzon, L. Beverina, R. Viswanatha, V. I. Klimov, S. Brovelli, *Nat. Photon.* **2014**, 8, 392; b) F. Meinardi, H. McDaniel, F. Carulli, A. Colombo, K. A. Velizhanin, N. S. Makarov, R. Simonutti, V. I. Klimov, S. Brovelli, *Nat. Nanotech.* **2015**, 10, 878; c) Y. Zhou, D. Benetti, Z. Fan, H. Zhao, D. Ma, A. O. Govorov, A. Vomiero, F. Rosei, *Adv. Energ. Mater.* **2016**, 6, 1501913; d) F. Meinardi, S. Ehrenberg, L. Dhamo, F. Carulli, M. Mauri, F. Bruni, R. Simonutti, U. Kortshagen, S. Brovelli, *Nat. Photon.* **2017**, 11, 177; e) K. Wu, H. Li, V. I. Klimov, *Nat. Photon.* **2018**, 12, 105; f) Z. Li, A. Johnston, M. Wei, M. I. Saidaminov, J. Martins de Pina, X. Zheng, J. Liu, Y. Liu, O. M. Bakr, E. H. Sargent, *Joule* **2020**, 4, 631; g) X. Liu, D. Benetti, J. Liu, L. Jin, F. Rosei, *Nano Energy* **2023**, 111, 108438; h) G. Liu, R. Mazzaro, Y. Wang, H. Zhao, A. Vomiero, *Nano Energy* **2019**, 60, 119.

[8] a) A. Anand, M. L. Zaffalon, G. Gariano, A. Camellini, M. Gandini, R. Brescia, C. Capitani, F. Bruni, V. Pinchetti, M. Zavelani-Rossi, F. Meinardi, S. A. Crooker, S. Brovelli, *Adv. Funct. Mater.* **2020**, 30; b) K. E. Knowles, T. B. Kilburn, D. G. Alzate, S. McDowall, D. Gamelin, *Chem. Commun.* **2015**, 51, 9129.

[9] C. Yang, H. A. Atwater, M. A. Baldo, D. Baran, C. J. Barile, M. C. Barr, M. Bates, M. G. Bawendi, M. R. Bergren, B. Borhan, C. J. Brabec, S. Brovelli, V. Bulović, P. Ceroni, M. G. Debije, J.-M. Delgado-Sanchez, W.-J. Dong, P. M. Duxbury, R. C. Evans, S. R. Forrest, D. R. Gamelin, N. C. Giebink, X. Gong, G. Griffini, F. Guo, C. K. Herrera, A. W. Y. Ho-Baillie, R. J. Holmes, S.-K. Hong, T. Kirchartz, B. G. Levine, H. Li, Y. Li, D. Liu, M. A. Loi, C. K. Luscombe, N. S. Makarov, F. Mateen, R. Mazzaro, H. McDaniel, M. D. McGehee, F. Meinardi, A. Menéndez-Velázquez, J. Min, D. B. Mitzi, M. Moemeni, J. H. Moon, A. Nattestad, M. K. Nazeeruddin, A. F. Nogueira, U. W.

Paetzold, D. L. Patrick, A. Pucci, B. P. Rand, E. Reichmanis, B. S. Richards, J. Roncali, F. Rosei, T. W. Schmidt, F. So, C.-C. Tu, A. Vahdani, W. G. J. H. M. van Sark, R. Verduzco, A. Vomiero, W. W. H. Wong, K. Wu, H.-L. Yip, X. Zhang, H. Zhao, R. R. Lunt, *Joule* **2022**, 6, 8.

[10] a) M. Siripurapu, F. Meinardi, S. Brovelli, F. Carulli, *ACS Photonics* **2023**, 10, 2987; b) W. van Sark, P. Moraitis, C. Aalberts, M. Drent, T. Grasso, Y. L'Ortije, M. Visschers, M. Westra, R. Plas, W. Planje, *Solar RRL* **2017**, 1, 1600015; c) T. A. de Bruin, R. Terricabres-Polo, A. Kaul, N. K. Zawacka, P. T. Prins, T. F. J. Gietema, A. C. de Waal, D. K. G. de Boer, D. A. M. Vanmaekelbergh, P. Leblans, S. Verkuilen, Z. Hens, C. de Mello Donega, W. G. J. H. M. van Sark, *Solar RRL* **2023**, 7, 2201121; d) N. S. Makarov, D. Korus, D. Freppon, K. Ramasamy, D. W. Houck, A. Velarde, A. Parameswar, M. R. Bergren, H. McDaniel, *ACS Appl. Mater. Interfaces* **2022**, 14, 29679; e) N. Aste, L. C. Tagliabue, C. Del Pero, D. Testa, R. Fusco, *Renew. Energy* **2015**, 76, 330.

[11] a) M. Kanellis, M. M. de Jong, L. Slooff, M. G. Debije, *Renew. Energy* **2017**, 103, 647; b) M. G. Debije, C. Tzikas, V. A. Rajkumar, M. M. de Jong, *Renew. Energy* **2017**, 113, 1288; c) M. G. Debije, C. Tzikas, M. M. de Jong, M. Kanellis, L. H. Slooff, *Renew. Energy* **2018**, 116, 335; d) Á. Bognár, S. Kusnadi, L. H. Slooff, C. Tzikas, R. C. G. M. Loonen, M. M. de Jong, J. L. M. Hensen, M. G. Debije, *Renew. Energy* **2020**, 151, 1141.

[12] X. You, C.-X. Wang, J. Huang, X. Gao, Z. Zhang, M. Wang, Y. Huang, C. Zhang, Y. Jiang, J. Wang, M. Zhu, B. Sheng, D. Wang, Z. Pan, P. Zhu, Y. Yang, Z. Liu, P. Zhang, X. Tao, S. Li, Z. Chen, X. Ma, C.-L. I, S. Han, K. Li, C. Pan, Z. Zheng, L. Hanzo, X. Shen, Y. J. Guo, Z. Ding, H. Haas, W. Tong, P. Zhu, G. Yang, J. Wang, E. G. Larsson, H. Q. Ngo, W. Hong, H. Wang, D. Hou, J. Chen, Z. Chen, Z. Hao, G. Y. Li, R. Tafazolli, Y. Gao, H. V. Poor, G. P. Fettweis, Y.-C. Liang, *Science China Information Sciences* **2020**, 64, 110301.

[13] M. A. Umair, M. Seminara, M. Meucci, M. Fattori, F. Bruni, S. Brovelli, F. Meinardi, J. Catani, *Laser & Photonics Reviews* **2023**, 17, 2200575.

[14] H. Haas, L. Yin, Y. Wang, C. Chen, *Journal of Lightwave Technology* **2016**, 34, 1533.

[15] H. G. Sandalidis, A. Vavoulas, T. A. Tsiftsis, N. Vaiopoulos, *Appl. Opt.* **2017**, 56, 3421.

[16] European Standard, *EN 410: 2011 - Glass in building - Determination of luminous and solar characteristics of glazing* **2011**.

[17] International Standard, *IEC 61215-1-1:2021 RLV, Terrestrial photovoltaic (PV) modules - Design qualification and type approval - Part 1-1: Special requirements for testing of crystalline silicon photovoltaic (PV) modules* **2021**.

[18] European Standard, *EN 1279-2:2018, Glass in building - Insulating glass units - Part 2: Long term test method and requirements for moisture penetration* **2018**.

[19] ISO Standard, *ISO 10077-1:2107, Thermal performance of windows, doors and shutters - Calculation of thermal transmittance)* **2017**.

[20] ISO Standard, *ISO 12631:2017, Thermal performance of curtain walling - Calculation of thermal transmittance* **2017**.

[21] A. N. S. Institute, *ASTM C1371-04a, Standard Test Method for Determination of Emittance of Materials Near Room Temperature Using Portable Emissometers* **2017**.

[22] ISO Standard, *ISO 12543-2:2021. Glass in building - Laminated glass and laminated safety glass - Part 2: Laminated safety glass* **2021**.

*Supporting Information*

**Methods**

**Materials**. Copper(I) iodide (CuI, purum, ≥99.5%), indium(III) acetate (In(OAc)$_3$, 99.99%) and 1-dodecanethiol (DDT, ≥98%) , sodium myristate (≥99%), selenium powder-100 mesh (99.99%), oleic acid, OA (≥90%), cadmium nitrate tetrahydrate (≥98%), copper(II) chloride dihydrate (99.999%) 1-octadecene, ODE (≥90%) were purchased from Sigma-Aldrich. Hexane (chromasolv, ≥97%), acetone (puriss. ≥99%), ethanol (puriss. ≥99%), methanol (puriss. ≥99%) were purchased from Honeywell *Riedel-de-Haën*. All the chemicals were used without further purification. Methylmethacrylate (MMA, 99%, Aldrich), purified with basic activated alumina (Sigma-Aldrich), was used as a monomer for the preparation of polymeric nanocomposites. 2,2′-Azobis(2-methylpropionitrile) (AIBN, 98%, Aldrich) and lauroyl peroxide (98%, Aldrich) were used as initiators without purification.

**Synthesis of CuInS$_2$/ZnS core/shell QDs.** The synthesis of CuInS$_2$ QDs was performed following a heat-up procedure. A mixture of CuI (8.0 mmol) and In(OAc)$_3$ (8.0 mmol) in 100 mL DDT was loaded into a 3-necked flask and degassed under reduced pressure at 60°C for 1 h. The reaction mixture was heated up to 220°C in nitrogen atmosphere and stirred for 30 min. The resulting dark red solution was then cooled to room temperature to quench the reaction. Without any further purification, the ZnS shell was grown on the CuInS$_2$ core QDs in the same reaction vessel. The shell precursors solution was prepared separately in a three-neck flask before injection. Zn(St)$_2$ (40.0 mmol) was added to a solution of 50 mL OA, 50 mL DDT and 100 mL ODE. The shell precursor solution was degassed for 30 min. at room temperature and then at 100°C for 1h, until complete dissolution of Zn(St)$_2$ and finally filled with nitrogen. After heating the core solution to 220°C the shell precursor solution was continuously injected with a syringe pump for 4 hours. The solution was then cooled to room temperature to quench the reaction Finally the crude CuInS$_2$/ZnS QDs solution was washed with a 1:2 hexane:ethanol solution and centrifuged at 3000 rpm for 5 minutes for three times.

**Fabrication of the LSC waveguide.** The synthesis of the PMMA/CuInS$_2$-ZnS QDs nanocomposite was carried out via industrial grade bulk polymerization of methyl methacrylate (MMA) using the so called "syrup" method. A batch of PMMA prepolymer (or syrup) consisting of a mixture of monomer and partially reacted oligomeric chains was prepared by adding 100ppm of 2,2'-azobis(2,4-dimethylpentanenitrile) (Vazo 52 by Chemours) and 1g of 1-Dodecanethiol (Merk) 8 Kg of MMA. The mixture was heated at 70°C and the reaction was for 28 min. Then, the syrup was quenched at 23°C and stored at 4°C in a fridge. 1g of CuInS$_2$-ZnS QDs was dispersed in 10 ml of hexane together with 8.8g of a commercial surfactant. The solution was kept under magnetic stirring and heated up to 50°C to ease the dispersion of the surfactant. The solution was added to 1kg of methyl methacrylate and the mixture was stirred for 5 min. Then, 800 mg of lauroyl peroxide were added and the solution was stirred for another minute. 1 kg of syrup at room temperature was slowly added to the MMA solution under continuous stirring and the resin was stirred for 15 minutes. The resin was casted in a mold consisting of 2 glass panes sealed with a PVC gasket. The mold was placed in a water bath at 52°C for 20h and then the temperature was increased to 92°C for 24h. The waveguide was laser cut to size and the edges were polished with a diamond blade.

**Characterization of the LSC waveguide.** The amount of residual monomer in the PMMA composites was extracted from the $^1$H Nuclear Magnetic Resonance (NMR) spectra, recorded on samples dissolved in deuterated chloroform by using an Avance 500 NMR spectrometer (Bruker). Tetramethylsilane was used as the internal standard. The glass transition temperature of the PMMA matrix was measured by Molecular weights and molecular weight distributions of PMMA matrices were determined by Gel Permeation Chromatography (GPC) using a WATERS 1515 isocratic equipped with a HPLC Pump, WATERS 2414 refractive index detector, four Styragel columns (HR2, HR3, HR4 and HR5 in the effective molecular weight range of 500–20 000, 500–30 000, 50 000–600 000 and 50 000–4 000 000 respectively) with tetrahydrofuran (THF) as the eluent at a flow rate of 1.0 ml min$^{-1}$. The GPC system was calibrated with standard polystyrene from Sigma-Aldrich. GPC samples were prepared by dissolution in THF. The solution was stirred at 80°C under reflux for 24h. The QDs were precipitated in THF and removed by centrifugation (6000 RPM for 15 minutes). The supernatant made of the polymeric matrix dissolved in the eluent was filtered with a hydrophobic PTFE membrane (pore size 0.2 μm) and measured.

**Assembly of the LSC-IGU.** LSC panels were laser cut to their final size. Edge polishing of the borders was performed to remove residual roughness. Customised 10 cm × 1 cm c-Si mini-modules, consisting of nine 1 cm$^2$ custom cells by NingboGZX PV Technology Co.Ltd mounted in series, were connected in parallel to form continuous PV strips of electrically independent modules of the same length as the LSC edge. The PV strips were then coupled to the LSC waveguide by in-situ polymerisation of a polyacrylate binder, which ensured mechanical robustness and a continuous refractive index across the LSC/PV cell interface. Then, the edges of the device (about 1.5 cm) were encapsulated in metal cable trays specially designed with a thin notch and a compressible filler to ensure minimal physical contact with the LSC waveguide - to minimise optical losses - and to compensate for the different thermal expansion of PMMA with respect to the borosilicate glass forming the external plates. Finally, the LSC was integrated into an IGU using standard assembly methods. It consists first in the formation of a sandwich made by a Pilkington OptiView™ OW 4 mm thick low iron tempered glass (external glass), a 12 mm thick aluminum spacer, the LSC, a 12 mm thick aluminum spacer and a Pilkington Optilam™ Therm S3 6.8 mm laminated glass with a low-emission coating (internal glass). The aluminum spacers were previously filled by synthetic zeolite desiccant Siliporite NK30 and laterally covered by polyisobutylene Fenzi Butylver. Then the sandwich was pressed and finally sealed by Dowsil 3363 silicone sealant.

**Characterization of LSC-IGUs.** To check the temporal stability of the LSC IGU across one year, a resistance temperature detector inserted within the IGU was continuously monitored, while a current sensor was installed to measure the photocurrent generated by the PV system in the open-circuit configuration. UV preconditioning was performed according to specifications of IEC 61215-1:2021 regulation[38], in a test chamber keeping temperature in the range (60 ± 5) °C. The irradiance in the band between 280 nm and 400 nm was checked to be below 250 W/m$^2$, with a uniformity of ± 15 %, no appreciable irradiance below 280 nm and with an irradiance between 3% and 10% in the range between 280 nm and 320 nm. The two samples were kept under irradiance for 228 hours and 900 hours respectively, providing a total irradiation of 15.1 kWh/m$^2$ and 60.1 kWh/m$^2$. Thermal cycling tests were performed according to specifications of IEC 61215-1:2021 regulation[38], and consisted of 50 full cycles between (-40 ± 2)°C and (85 ± 2) °C starting from an initial temperature of 25°C, with a maximum heating/cooling rate of 100°C/h and minimum dwell times of 10 minutes. Humidity freeze tests were performed according to specifications of IEC 61215-1:2021 regulation[38], and consisted in ten full cycles consisting of 20 hours at (85 ± 2) °C and (85 ± 5) % relative humidity (RH) followed by 30 minutes at (-40 ± 2)°C with a cooling rate of 100°C/h (between +85°C and 0°C) and 200°C/h (between 0°C and -40°C). Damp heat tests were performed according to specifications of IEC 61215-1:2021 regulation[38], and consisted of heating the sample in a climate chamber starting from room temperature to (85 ± 2) °C with a maximum heating of rate 100°C/h, and keeping for 200 hours at this temperature and (85 ± 5) % RH. The PV characteristics of the LSC-IGUs before and after aging tests were assessed using a calibrated flash test with simulated sunlight, which is commonly used to measure the output power compliance of a solar PV module, in accordance with the IEC 61215-1:2021 regulation[38]. The power reduction of the module after ageing should be below 5%. The moisture penetration moisture penetration index $I$ has been evaluated according to EN 1279-2:2018[39]. It consists of evaluating the percentage of water content (indicated as $W$) in molecular sieves placed inside the sealed chamber of IGU prototypes before and after an ageing period. The complete ageing prescribed by EN 1279-2:2018[39] consists of 11 weeks, specifically consisting of 56 thermal cycles between (-18 ± 2) °C and (53 ± 2) °C at RH>95% for a total period of 4 weeks, followed by 7 weeks at a constant temperature of (58 ± 2) °C and RH>95%. Since these are destructive testing, in order to perform these measurements 35 identical LSC-IGUs were produced to monitor them at different stages of the ramp. The $W$-value is measured at the beginning ($W_i$) and at the end ($W_f$) of the test. Given the moisture adsorption capacity of dessiccant $T_c$ it is possible to determine the moisture penetration index $I = (W_f - W_i )/( W_c - W_f)$. The test is passed if, at the end of the full ramp, the average $I$-value between five different prototypes is $I_{AV}$<20% and no prototype has $I$>25%. The luminous transmittance of our QD-LSC waveguides has been measured according to the EN 410:2011[40]. The regular spectral transmittance at normal incidence and spectral reflectance at an angle of incidence of 5° in the 300-2500 nm range were measured using a dual-beam spectrophotometer with an aperture angle of 2°, providing near-parallel and near-normal incident radiation. A 6 mm thick Suprasil W ultra-pure silica glass was used as a reference for reflectance measurements. The photometric accuracy of the spectrophotometer in the visible range was checked using certified filters (SRM 930D). The light transmittance, light reflectance, direct solar transmittance, direct solar reflectance and solar factor of the glazing were then determined according to the calculation procedures specified in EN 410:2011[40]. The hemispherical surface emissivity, $E$, of the QD-LSC at room temperature was assessed in accordance with ASTM C1371-04a protocol[43], interfacing a calibrated emissometer with a scaling digital voltmeter. The solar radiation resistance of the spectral characteristics was evaluated according to the UNI EN ISO 12543-4:2022 standard[44] by exposing the samples under test to 16 OSRAM ULTRAVITALUX 300 W lamps for 2000 hours at 45±5 °C, as specified in paragraph 7.3.1 of the standard (Method A). The light transmittance before and after irradiation was measured in accordance with the EN 410:2011[40]. The requirement of UNI EN ISO 12543-4:2022 standard[44] is that the transmittance of three samples irradiated should not vary by more than ± 2% of their absolute initial value.

**Thermal insulation calculations.** The thermal transmittance of our complete LSC-IGU for window applications has been be calculated following the protocols described in ISO 10077-1:2017 standard[41,42] and described in detail in the dedicated section below.

| Transparency (%) | 90 | 80 | 70 | 60 | 50 | 40 | 30 | 20 | 10 |
|---|---|---|---|---|---|---|---|---|---|
| CRI | 100 | 100 | 98 | 98 | 98 | 97 | 96 | 93 | 89 |
| UNI 10380 Class | 1A | 1A | 1A | 1A | 1A | 1A | 1A | 1A | 1B |

**Table S1.** Color rendering index of LSC IGUs and respective class according to UNI 10380 standard

**Thermal insulation calculations.**

Thermal transmittance of complete IGU

The thermal transmittance of our complete LSC-IGU for window applications ($U_w$) has been be calculated following the protocols described in ISO 10077-1:2017 standard[41,42] as:

$$U_w = \frac{A_g U_g + A_f U_f + l_g \Psi_g}{A_w}$$

where $U_g$ is the glazing thermal transmittance, $U_f$ is the framing thermal transmittance and $\Psi_g$ is the frame-glass junction thermal transmittance (parameters calculated as in following sections). On the other hand, $A_g$ is the glazing area, $A_f$ is the framing area, $l_g$ is the framing length and $A_w$ is the window area. For the calculations of the parameters shown in **Figure 3g**, we used the geometric parameters reported in Table S2.

| Parameter | Value |
|---|---|
| Window dimension 1 ($L_1$) | 1 m |
| Window dimension 2 ($L_2$) | 1 m |
| Window area ($A_w$) | 1 m$^2$ |
| Framing length (perimeter) | 4 m |
| Framing width | 0.11 m |
| Glazing area ($A_g$) | 0.79 m$^2$ |
| Framing area ($A_f$) | 0.21 m$^2$ |

**Table S2.** Geometric parameters used for the estimation of $U_w$.

Glazing thermal transmittance

The glazing thermal transmittance $U_g$ is calculated according to the iteration procedure described in EN 673:2011, annex A[RR1]. For systems like those described in this work (which consist of IGUs with 2 gas spaces) the procedure imposes to set the initial values of the temperature differences $\Delta T_1$ (between first and second glass panes) and $\Delta T_2$ (between second and third glass panes) as $\Delta T_1 = \Delta T_2 = 7.5$ K. From these values, one can determine the Nusselt number $Nu_k$ of each gas space $k$ as follows:

$$\mathrm{Nu}_k = \max\left\{1;\ A\left(\frac{9.81\, s_k^3\, \rho_k^2\, c_k}{\mu_k\, \lambda_k\, T_{M,k}}\, \Delta T_k\right)^n\right\} \qquad (***)$$

where $n=0.38$, $A=0.035$ (for vertical glazing), $s_k$ is the thickness of the $k$-th gas space, $\rho_k$ is the gas density, $c_k$ the specific heat capacity, $\mu_k$ the dynamic viscosity, $\lambda_k$ the thermal conductivity and $T_{M,k}$ is the mean temperature of the $k$-th gas space (conventionally, $T_{M,k} = 283$ K). The gas conductance $h_{g,k}$ and the radiation conductance $h_r$ of each space $k$ are determined as:

$$h_{g,k} = \frac{\mathrm{Nu}_k\, \lambda_k}{s_k} \qquad h_{r,k} = 4\sigma\left(\frac{1}{\varepsilon_{1,k}} + \frac{1}{\varepsilon_{2,k}} - 1\right)^{-1} T_{M,k}^3$$

where $\sigma$ is the Stefan-Boltzmann's constant, while $\varepsilon_{1,k}$ and $\varepsilon_{2,k}$ are the corrected emissivities of the surface bounding the gas space $k$. These quantities allow to determine the heat transfer $h_{s,k}$ of each space $k$ as $h_{s,k} = h_{g,k} + h_{r,k}$. This allows to determine the new $\Delta T_k$ values as:

$$\Delta T_k = 15 \frac{(1/h_{s,k})}{\sum_{k=1}^{2}(1/h_{s,k})}$$

which can then be used for following iterations starting from Eq. (\*\*\*). Once that the values of $h_{s,k}$ converged, the total thermal conductance of the glazing $h_t$ is calculated as:

$$h_t = \left(\sum_{k=1}^{2} \frac{1}{h_{s,k}} + \sum_{j=1}^{3} d_j\, r_j\right)^{-1}$$

where $d_j$ is the thickness of each material layer, while $r_j$ is its thermal resistivity. The glazing thermal transmittance $U_g$ is then finally calculated as:

$$U_g = \left(\frac{1}{h_t} + \frac{1}{h_i} + \frac{1}{h_e}\right)^{-1}$$

where $h_i$ and $h_e$ are the heat transfer coefficients of the internal and the external glasses. The parameters used in this work for the calculation of $U_g$ are shown in Table S3.

| Parameter | Value | Source |
|---|---|---|
| Argon density $\rho_1 = \rho_2$ | 1.699 kg/m$^3$ | 46 |
| Argon specific heat capacity $c_1 = c_2$ | 0.519 x 10$^3$ J/(kg K) | 46 |
| Argon dynamic viscosity $\mu_1 = \mu_2$ | 2.164 x 10$^{-5}$ kg/(m s) | 46 |
| Argon thermal conductivity $\lambda_1 = \lambda_2$ | 1.684 x 10$^{-2}$ W/(mK) | 46 |
| Uncoated glass thermal emissivity $\varepsilon$ | 0.837 | 46 |
| Coated glass thermal emissivity $\varepsilon$ | 0.037 | 46 |
| PMMA thermal emissivity $\varepsilon$ | 0.91 | Meas. as in [43] |
| Glass thermal resistivity $r$ | 1.0 m K / W | 46 |
| PMMA thermal resistivity $r$ | 5.6 m K / W | https://www.goodfellow.com/us/en-us/displayitemdetails/p/me30-sh-000100/polymethylmethacrylate-sheet |
| Heat transfer coefficient of the external glass $h_e$ (uncoated soda lime glass surface) | 25 W / m$^2$ K | 46 |

| Heat transfer coefficient of the internal glass $h_i$ (uncoated soda lime glass surface) | 7.7 W / m² K | 46 |

**Table S3.** Parameters used for the calculation of $U_g$.

Framing thermal transmittance

The framing thermal transmittance $U_f$ is numerically calculated using the software tool (Heat Transfer Module of the FEM software Comsol Multiphysics) according to the method described in the standard ISO 10077-2:2017 annex F[41,42]. According to this method, the IGU is replaced by an insulation panel of identical thickness, whose thermal conductivity is set to $\lambda$ = 0.035 W/(m K) and whose width $b_p$ is set to 190 mm. The inner face of the pane is set to 20 °C, while the external side is set to 0 °C. The simulation allows to determine the two-dimensional thermal conductance $L_f^{2D}$ and the thermal transmittance of the central area of the pane $U_p$, from which the framing thermal transmittance $U_f$ can be calculated as:

$$U_f = \frac{L_f^{2D} - U_p \, b_p}{b_f}$$

being $b_f$ = 110 mm the frame profile width (in our situation, we used the wood frame profile by IFT Rosenheim (see ift Guideline WA-08engl/3 – "Thermally improved spacer - Part 1: Determination of the representative Psi value for window frame profiles")). The materials parameters used in the simulation are reported in Table S4.

| Material | Density [kg/m³] | Thermal conductivity [W/(m K)] | Heat capacity at const. temp. [J/(kg K)] |
|---|---|---|---|
| Wood | 500 | 0.13 | 1600 |
| Insulating EDPM | 1150 | 0.25 | 1000 |
| Glass | 2500 | 1 | 750 |
| Polysulfide | 1700 | 0.40 | 1000 |
| Aluminum | 2800 | 160 | 880 |
| Stainless steel | 7500 | 15 | 502 |
| Silica gel | 720 | 0.13 | 1000 |
| PMMA | 1180 | 0.16 | 1.42 |

**Table S4.** Parameters used in the simulations to calculate $U_f$ and $\Psi_g$ (parameters from the libraries of software tool Heat Transfer Module of the FEM software Comsol Multiphysics).

Frame-glass junction thermal transmittance

The frame-glass junction thermal transmittance $\Psi_g$ is calculated according to the method described in the standard ISO 10077-2:2017 annex F[41,42], according to the software tool. This is done analogously to the method used for the calculation of $U_f$, but considering the presence of the real glazing despite of the insulation panel. In this case, it is possible to calculate the two-dimensional thermal conductance $L_\Psi^{2D}$ and the thermal transmittance of the central area of the glazing $U_g^*$, from which the frame-glass junction thermal transmittance $\Psi_g$ is calculated as follows:

$$\Psi_g = L_\Psi^{2D} - U_f\, b_f - U_g^*\, b_g$$

being $b_g$ the width of the glazing width used in the simulation (in this study 190 mm). The material parameters used in the simulation are reported in Table S3.

**LSC-IGU for wireless communications**

**Characterization of LSG-IGU intrinsic bandwidth for VLC communication and bit rate selection.** To choose the most suitable bitrate for VLC links in realistic settings, the intrinsic LSC bandwidth (BW) as well as the BW of the complete system are evaluated (**Figure S1a**). Error bars correspond to standard deviation over 3 to 5 repeated runs. For the first test we use a low-power, fast 405 nm LED (Thorlabs, LED405E) driven by an Arbitrary Waveform Generator (AWG) (Siglent, SDG2042X), under small modulation depth conditions and a fast APD as detector (Thorlabs, APD430A2) positioned in the middle point of one of the slab's edges. The LED is positioned in contact with the central point of the slab's surface, and the optical signal collected on the edge is recorded by varying the frequency on the AWG. Experimental points are expressed in dB, considering the maximum of the signal as reference (blue symbols in **Figure S1a**). As expected[45] the LSC BW behaves as a first order RC filter, with a measured cutoff frequency (-3dB) around 300kHz. We note that the intrinsic BW (12 MHz) of LED source and of detector (400 MHz), being much higher than that of the LSC, do not significantly affect the measurement of the LSC BW. We then evaluate the global BW of our VLC system in the proposed scenario, using the spotlight LED as light source and the S3588-09 rectangular PD as detector as in **Figure 4a**, LSC is positioned in Line of Sight (LoS) at a distance of 5 meters from the optical source in order to reproduce a fairly uniform illumination on the LSC. Experimental measurement of the complete BW is shown in **Figure S1a** (green symbols). The 3 kHz low-frequency cut-off (-3dB) is introduced by both the current modulator that drives the spotlights and the AC decoupling of the detector. Indeed, the PD input stage has been modified in order to filter out the DC component of the optical signal (cut off frequency ≈ 2-3 kHz), essential to ensure efficient communication especially under high solar irradiance conditions[35]. Considering the results of total BW obtained, we select a bit rate of 500 kb/s OOK (On Off Keying) with Manchester encoding for data transmission. Higher bit rates can be achieved with appropriate pre- and post-equalization techniques and using higher-order modulation schemes. We also note that commercially available high-power spotlights often exploit COB-LED sources, typically featuring strong BW limitations as a consequence of the large extension of the substrate, inducing large parasitic capacitances. Larger BWs are expected when using LED arrays as TX sources.

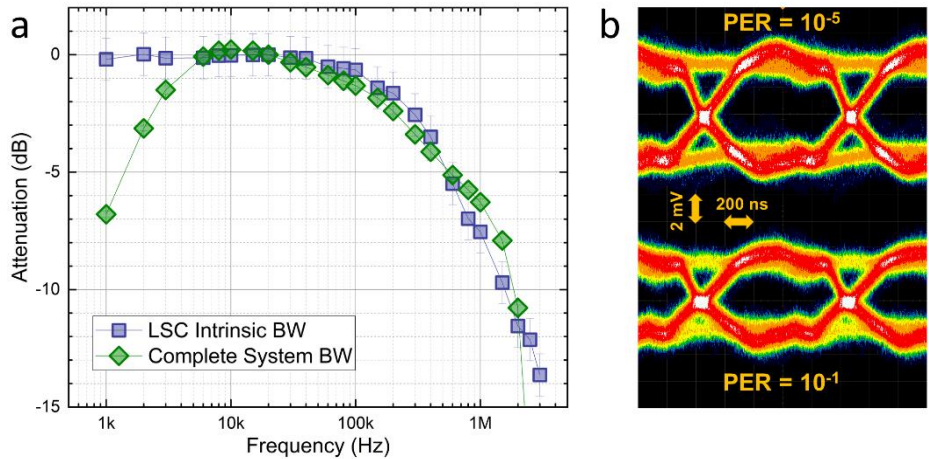

**Figure S1. QD-LSC as optical antennas for VLC Li-Fi communication. a.** Bandwidth characterization: Blue symbols correspond to LSC intrinsic BW. Green symbols correspond to the BW of the complete VLC system (light source + LSC-based receiver). **b.** Acquired Eye-Diagrams for two SNR levels corresponding to PER = $10^{-1}$ and PER = $10^{-5}$.

**Eye diagrams.** As a further analysis, we also report eye-diagrams (**Figure S1b**) on the received signal for different conditions (PER = $10^{-1}$ and PER $10^{-5}$ for 500 kb/s OOK Manchester), recorded using a digital oscilloscope (Keysight, InfiniiVision DSOX6004A). Eye Diagrams were acquired directly on the photodiode output before the discrimination stage. The distortion of the signal is due to a combined effect of BW limits in both TX and RX stages, as well as to a non-perfect impedance matching between current modulator and light source. These impairments could be easily mitigated by means of pre-equalization techniques in future implementations.